\documentclass[a4paper,12pt]{article}
\usepackage{amsmath}
\makeatletter
\def\thebibliography#1{\section{\refname\@mkboth
 {\uppercase{\refname}}{\uppercase{\refname}}}\list
{\@biblabel{\arabic{enumiv}}}{\settowidth\labelwidth{\@biblabel{#1}}
\leftmargin\labelwidth
  \advance\leftmargin\labelsep
   \usecounter{enumiv}    \let\p@enumiv\@empty
  \def\theenumiv{\arabic{enumiv}}}    \def\newblock{\hskip .11em plus.33em
minus.07em}    \sloppy\clubpenalty4000\widowpenalty4000
  \sfcode`\.=1000\relax}
\makeatother

\begin{document}

\title{Heun and Mathieu functions as solutions of the Dirac equation}
\author{T.Birkandan$^{*}$ and M. Horta\c{c}su$^{*}$$^{1}$\\
%EndAName
{\small $^{*}${Department of Physics, Istanbul Technical University,
Istanbul, Turkey}. }\\
}
\date{\today}
\maketitle

\begin{abstract}
\noindent
We give examples of where the Heun function exists as
solutions of wave equations encountered in general relativity. While
the Dirac equation written in the background of Nutku helicoid
metric yields Mathieu functions as its solutions in four spacetime
dimensions, the trivial generalization to five dimensions results in
the double confluent Heun function. We reduce this solution to the
Mathieu function with some transformations. We must apply
Atiyah-Patodi-Singer spectral boundary conditions to this system
since the metric has a singularity at the origin.
\end{abstract}

\bigskip\bigskip

\bigskip\bigskip\bigskip\bigskip\bigskip\bigskip\bigskip\bigskip\bigskip\bigskip\bigskip\bigskip\bigskip\bigskip\bigskip\bigskip\bigskip\bigskip\bigskip\bigskip\bigskip
\footnotetext[1]{E-mail addresses: hortacsu@itu.edu.tr,
birkandant@itu.edu.tr} \pagebreak

\maketitle

\section{Introduction}

\noindent Most of the theoretical physics known today is described
by a rather few number of differential equations. If we study only
linear problems, the different special cases of the hypergeometric
or the confluent hypergeometric equations often suffice to analyze
scores of different phenomena. Both of these equations have simple
recursive relations between consecutive coefficients when series
solutions are sought. They also have simple integral
representations. If the problem is nonlinear then one can often use
one of the forms of the Painlev\'{e} equation. In the linear case,
sometimes it is necessary to work with more complicated equations.
Such an equation is the Heun equation, whose special forms we
encounter as the Mathieu, Lam\'{e}, Coulomb spheroidal equations.
These equations do not have two way recursion relations. There are
several examples in the literature for this kind of equations
\cite{ronveaux} \cite{siopsis} \cite{fiziev}.

\noindent We "encountered" this equation when we tried to solve the
scalar and Dirac equations in the background of the Nutku helicoid
instanton \cite{nutku} \cite{alievnutku} \cite{alievetal}
\cite{birkandanhortacsu1} \cite{birkandanhortacsu2}.

\section{Dirac Equation}

\noindent The original four dimensional Nutku helicoid metric is
given as
\begin{eqnarray}
ds_{4}^{2}=\frac{a^{2}}{2}\sinh 2x(dx^{2}+d\theta ^{2})   \nonumber
+\frac{2}{\sinh 2x}[(\sinh ^{2}x+\sin ^{2}\theta )dy^{2} \\
-\sin2\theta dydz+(\sinh ^{2}x+\cos ^{2}\theta )dz^{2}]
\end{eqnarray}

\noindent where $0<x<\infty $, $0\leq \theta \leq 2\pi $, $y$ and
$z$ are along the Killing directions and will be taken to be
periodic coordinates on a 2-torus \cite{alievetal}. This is an
example of a multi-center metric.

\noindent Sucu et al studied the solutions of the Dirac equation in
the background of this metric \cite{sucuunal}. In their paper they
study the spinor field in the background of the Nutku helicoid
instanton. They obtain an exact solution, which, however, can be
expanded in terms of Mathieu functions \cite{chaosetal}
\cite{birkandanhortacsu1}.

\noindent Here we study the same problem in five dimensions. The
addition of the time component to the previous metric gives
yields, $ds^{2}=-dt^{2}+ds_{4}^{2}$, resulting in the massless
Dirac equation as:
\begin{equation}
\frac{\sqrt{2}}{a\sqrt{\sinh 2x}}\{(\partial _{x}+i\partial _{\theta })\Psi
_{3}   \\
+a[\cos (\theta +ix)\partial _{y}+\sin (\theta +ix)\partial _{z}]\Psi _{4}+i%
\frac{a\sqrt{\sinh 2x}}{\sqrt{2}}\partial _{t}\Psi _{1}\}=0,
\end{equation}
\begin{equation}
\frac{\sqrt{2}}{a\sqrt{\sinh 2x}}\{(\partial _{x}-i\partial _{\theta })\Psi
_{4}   \\
-a[\cos (\theta -ix)\partial _{y}+\sin (\theta -ix)\partial _{z}]\Psi _{3}+i%
\frac{a\sqrt{\sinh 2x}}{\sqrt{2}}\partial _{t}\Psi _{2}\}=0,
\end{equation}
\begin{equation}
\frac{\sqrt{2}}{a\sqrt{\sinh 2x}}\{(\partial _{x}-i\partial _{\theta }+\coth
2x)\Psi _{1}   \\
-a[\cos (\theta +ix)\partial _{y}+\sin (\theta +ix)\partial _{z}]\Psi _{2}-i%
\frac{a\sqrt{\sinh 2x}}{\sqrt{2}}\partial _{t}\Psi _{3}\}=0,
\end{equation}
\begin{equation}
\frac{\sqrt{2}}{a\sqrt{\sinh 2x}}\{(\partial _{x}+i\partial _{\theta }+\coth
2x)\Psi _{2}   \\
+a[\cos (\theta -ix)\partial _{y}+\sin (\theta -ix)\partial _{z}]\Psi _{1}-i%
\frac{a\sqrt{\sinh 2x}}{\sqrt{2}}\partial _{t}\Psi _{4}\}=0.
\end{equation}

\noindent If we solve for $\Psi _{1}$ and $\Psi _{2}$ and replace them in
the latter equations, we get two equations which has only $\Psi _{3}$ and $%
\Psi _{4\text{ }}$ in them. If we take $\Psi
_{i}=e^{i(k_{t}t+k_{y}y+k_{z}z)}{\mathbf{\Psi _{i}}}(x,\theta )$,
the resulting equations read:

\begin{equation}
\left\{ \partial _{xx}+\partial _{\theta \theta }+\frac{a^{2}k^{2}}{2}%
\left\{ \cos [2(\theta +\phi )]-\cosh 2x\right\} +2a^{2}k_{t}^{2}\sinh
2x\right\} \bigskip \Psi _{3,4}=0 .
\end{equation}
\noindent If we assume that the result is expressed in the product
form $\Psi _{3}=T_{1}(x)T_{2}(\theta ),$ the angular part is
expressible in terms of Mathieu functions.
\begin{eqnarray}
T_{2}(\theta ) =Se\left[ \eta,-\frac{a^{2}k^{2}}{4},\arccos (\sqrt{\frac{%
1+\cos (\theta +\phi )}{2}})\right]  \nonumber \\
+So\left[ \eta,-\frac{a^{2}k^{2}}{4},\arccos (\sqrt{\frac{1+\cos
(\theta +\phi )}{2}})\right] .
\end{eqnarray}
\noindent Here $\eta$ is the separation constant and it is equal
to the square of an integer because of the periodicity of the
solution.

\noindent The equation for $T_{1}$ reads:

\begin{equation}
\left\{ \partial _{xx}-\frac{a^{2}k^{2}}{2}\cosh 2x+2a^{2}k_{t}^{2}\sinh
2x-\eta\right\} T_{1}=0
\end{equation}
\noindent The solution of this equation is expressed in terms of
double confluent Heun functions \cite{ronveaux}.

\begin{eqnarray}
T_{1}(x) &=&H_{D}\left[ 0,\frac{a^{2}k^{2}}{2}+\eta ,4a^{2}k_{t}^{2},\frac{%
a^{2}k^{2}}{2}-\eta ,\tanh x\right]  \nonumber \\
&&+H_{D}\left[ 0,\eta +\frac{a^{2}k^{2}}{2},4a^{2}k_{t}^{2},\frac{a^{2}k^{2}%
}{2}-\eta ,\tanh x\right] \\
&&\times \int \frac{-dx}{H_{D}\left[ 0,\eta +\frac{a^{2}k^{2}}{2}%
,4a^{2}k_{t}^{2},\frac{a^{2}k^{2}}{2}-\eta ,\tanh x\right] ^{2}}.  \nonumber
\end{eqnarray}

\noindent We only take the first function and discard the second solution.

\noindent There is an obstruction in odd Euclidean dimensions that
makes us use the Atiyah-Patodi-Singer (APS) spectral boundary
conditions \cite{atiyahetal}. These boundary conditions can also be
used in even Euclidean dimensions if we want to respect the charge
conjugation and the $\gamma^5$ symmetry \cite{hortacsuetal}
\cite{hortacsu} \cite{birkandanhortacsu2}.

\noindent Just to show the differences with the four dimensional solution,
we attempt to write this expression in terms of Mathieu functions. This can
be done after few transformations. We define $A=2a^{2}k_{t}^{2},B=-\eta ,C=-%
\frac{a^{2}k^{2}}{2},$and use the transformation $z=e^{-2x}.$Then the
differential operator is expressed as

\begin{equation}
\mathit{O}=4z^{2}\partial _{zz}+4z\partial _{z}+A^{\prime }z+B+C^{\prime }%
\frac{1}{z}.
\end{equation}%
\noindent Here $A^{\prime }=\frac{C-A}{2},C^{\prime }=\frac{C+A}{2}$. $\sqrt{\frac{%
C^{\prime }}{A^{\prime }}}u=z,$ $w=\frac{1}{2}(u+\frac{1}{u})$ and set $E=%
\sqrt{A^{\prime }C^{\prime }}$ we get,

\begin{equation}
\mathit{O}=(w^{2}-1)\partial _{ww}+w\partial _{w}+\frac{E}{2}w+\frac{B}{4}.
\end{equation}
\noindent The solution of this equation is expressible in terms of
Mathieu functions given as:

\begin{equation}
R(z)=Se(-B,E,\arccos \sqrt{\frac{w+1}{2}})+So(-B,E,\arccos \sqrt{\frac{w+1}{2%
}})
\end{equation}

\noindent Although both the radial and the angular part can be
written in terms of Mathieu functions, the constants are
different,modified by the presence of the new $-2a^2 k_{t^2}^2$
term, which makes the summation of these functions to form the
propagator quite difficult.

\section{Laplacian}

\noindent The Laplacian can be solved by separation of variables
method. The solution of the angular part can be expressed in terms
of Mathieu functions. The solution of the radial part is can be
written in terms of Double confluent Heun functions, which can be
reduced to the modified Mathieu function after performing similar
transformation as in the spinor case \cite{birkandanhortacsu2}.

\section{Conclusion}

\noindent Here we related solutions of the Dirac equation in the
background of the Nutku helicoid solution in five dimensions to
the Double confluent Heun function. The solution can be also
expressed in terms of the Mathieu function at the expense of using
a transformation. Often increasing the number of dimensions of the
manifold results in higher functions as solutions. Here we call a
function of a higher type if it has more singularities. In this
respect Heun function belongs to a higher form than the
hypergeometric function. APS Boundary conditions should be used in
five dimensional case. They can also be used in four dimensional
case for physical purposes. With work increasing on higher
dimensions, it should not be far when we will encounter Heun
functions more often in theoretical physics literature.

\noindent \linebreak \textbf{Acknowledgement}: We would like to thank to the
organizing committee of the Spanish Relativity Meeting 2007.

%%-----------------------------
%%      your bibliography
%%-----------------------------

\end{document}